\documentclass[twocolumn,amsmath,aps]{revtex4}
\usepackage{graphicx}

\newcommand{\nl}{\nonumber \\}

\newcommand{\be}{\begin{equation}}
\newcommand{\ee}{\end{equation}}
\newcommand{\bea}{\begin{eqnarray}}
\newcommand{\eea}{\end{eqnarray}}

\newcommand{\Eq}[1]{Eq.\,(\ref{#1})}
\newcommand{\Eqs}[1]{Eqs.\,(\ref{#1})}

\newcommand{\ra}{\rangle}
\newcommand{\dg}{\dagger}

\begin{document}


\title{\large Quantum measurement characteristics of
              double-dot single electron transistor}

\author{HuJun Jiao}
\email{jiaohujun@semi.ac.cn}
\affiliation{State Key Laboratory for
Superlattices and Microstructures,
         Institute of Semiconductors,
         Chinese Academy of Sciences, P.O.~Box 912, Beijing 100083, China}
\author{JunYan Luo}
\affiliation{State Key Laboratory for Superlattices and Microstructures,
         Institute of Semiconductors,
         Chinese Academy of Sciences, P.O.~Box 912, Beijing 100083, China}
\affiliation{Department of Chemistry, Hong Kong University of
             Science and Technology, Kowloon, Hong Kong}
\author{Xin-Qi Li}
\email{xqli@red.semi.ac.cn} \affiliation{State Key Laboratory for
Superlattices and Microstructures,
         Institute of Semiconductors,
         Chinese Academy of Sciences, P.O.~Box 912, Beijing 100083, China}

\date{\today}

\begin{abstract}
Owing to a few unique advantages,
double-dot single electron transistor has been proposed
as an alternative detector for charge states.
In this work, we present a further study for its signal-to-noise
property, based on a full analysis of the setup configuration
symmetry. It is found that the effectiveness of the double-dot
detector can approach that of an ideal detector, if the symmetric
capacitive coupling is taken into account. 
The quantum measurement efficiency is also analyzed, by comparing
the measurement time with the measurement-induced dephasing time.
\end{abstract}
\maketitle

\section*{Introduction}
Quantum measurement in solid-state mesoscopic systems has
attracted considerable interest in the past
years\cite{Dev00,Hsi01,Shn98,Sch01,Ale97,Gur96,Gur05}. Besides the
intensive theoretical work , experimental progresses are in
particular impressive
\cite{Buk98,Sch98,Mac01,van03,Elz04,Xia04,Gus06,Fuj06}. 
In these studies, two measurement devices were typically focused
on, i.e., the mesoscopic quantum point contact (QPC) and the
single electron transistor (SET).
Usually, the SET is restricted to the device with a single dot
embedded in between the source and drain electrodes.
Very recently, the double-dot (DD) SET has been proposed as an alternative
charge detector \cite{Gat02,Bre03,Bre05,Tan04,Ges06,Gur06}.
Compared to the single-dot detector, in addition to the obvious advantage of
weakening the requirement of very low temperature,
the DD detector may have other advantages such as:
(i) It can probe the rapid transitions
between electrostatically degenerate charge states \cite{Bre03}.
Experimentally, its radio-frequency operation has been demonstrated \cite{Bre05}.
(ii) DD detector is able to probe the entanglement of two qubits \cite{Tan04}.
(iii) Most importantly, DD detector has better immunity against noises \cite{Ges06}.

Owing to the added complexity of the DD detector, better
understanding of its measurement dynamics is of interest and seems
a timely work at this stage.
Very recently, this problem was studied by Gilad and Gurvitz
\cite{Gur06}. The key insight gained in their work is the {\it
symmetry} property of the setup configuration, which is revealed
in terms of the response current of the DD detector in both the
time and frequency domains. However, their analysis was based on
an {\it extremely asymmetric} capacitive coupling configuration,
which leads to a conclusion that the DD detector is a sensitive
detector, but {\it cannot} reach the signal-to-noise ratio of 4,
i.e., the value of an ideal QPC detector.

In this work, we present a further study for the signal-to-noise
property of the DD detector, based on a full analysis of the
capacitive coupling symmetry. 
In contrast with Ref.\ \onlinecite{Gur06}, we conclude that the DD
detector can approach the signal-to-noise ratio of an ideal QPC
detector, if the symmetric capacitive coupling setup is taken into
account. 
Moreover, we also analyze the {\it quantum measurement efficiency} of the
DD detector, by comparing the measurement time with the
measurement-induced dephasing time. It is found that, under the
setup configuration that results in the optimal signal-to-noise ratio, the
measurement efficiency cannot reach unity (i.e. the value of ideal QPC
detector). However, in principle, it can approach unity under
proper parametric conditions.

\begin{figure}[h]
\begin{center}
\includegraphics[width=8cm]{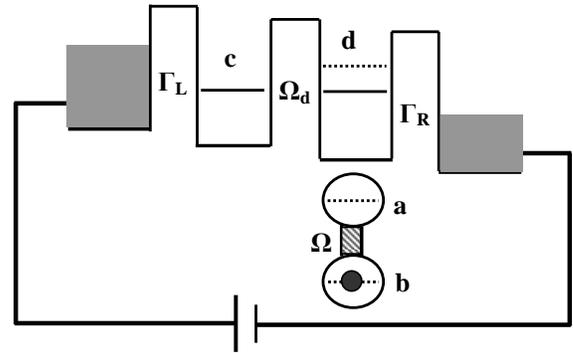}
\caption{Schematic setup of using the double-dot single electron
transistor to perform quantum measurement of a solid-state qubit.}
\end{center}
\end{figure}

\section*{Model Description}
As schematically shown in Fig.1, let us consider a charge qubit measured
by a mesoscopic transport device.
The charge qubit studied here is modeled by a pair of coupled
quantum dots with an extra electron in it,
while the detector is the proposed DD single electron transistor.
The entire system is described by the following Hamiltonian
\begin{subequations}
\begin{align}
H=&H_0+H^\prime\\
H_0=&H_s+\sum_k(\epsilon_k^Lc_k^\dag c_k+\epsilon_k^Rd_k^\dag
d_k)\label{Eq1}\\
H_s=&\sum_{i=a,b,c,d}E_ia_i^\dag a_i+\Omega(a_a^\dag a_b+a_b^\dag
a_a)+\Omega_d(a_c^\dag a_d+a_d^\dag a_c)  \nl
& +\sum_{i=a,b}\sum_{j=c,d} U_{ij}n_in_j+U_{cd}n_cn_d\\
H^\prime=&\sum_k(\Omega_k^La_c^\dag c_k+\Omega_k^Ra_d^\dag
 d_k+{\rm H.c.})\nl
\equiv  & a_c^\dag f_c+a_d^\dag f_d+ {\rm H.c.}
\end{align}
\end{subequations}
In these decomposed Hamiltonians, $a_a^\dag(a_a)$,
$a_b^\dag(a_b)$, $a_c^\dag (a_c)$, $a_d^\dag(a_d)$, $c_k^\dag
(c_k)$ and $d_k^\dag (d_k)$ are the electron
creation (annihilation) operators of the qubit,
detector's two central dots and the reservoirs.
In the following treatment, the tunneling Hamiltonian $H^{\prime}$
of the DD detector will be taken as perturbation.
The free Hamiltonian in the above, $H_0$, consists of
the detector's reservoirs, its central two dots, the qubit,
and the Coulomb interaction between them.

In this work, we assume that the DD detector works in the strong
Coulomb-blockade regime, i.e., there will be at most one more
electron occupied in the two dots.
Therefore, only the three DD states $|00\rangle$, $|10\rangle$,
and $|01\rangle$ are involved in the transport process.
Here, 0 and 1 stand for the vacant and occupied dot states,
while their ordering in ``$|\cdots\rangle$" is from the left to the right
dot states of the detector.
For the qubit, it has two logic states, i.e., the dot states $|a\ra$ and $|b\ra$.
For the sake of simplicity, we assume that each dot has only one bound state.
Intuitively, the measurement principle of the device under study is as follows:
if the qubit is in state $|b\ra$, the two states $|10\rangle$ and $|01\rangle$
of the DD detector is nearly energetically degenerate;
while the qubit is in state $|a\ra$, they will be in off-resonance,
due to the relatively stronger Coulomb interaction $U_{ad}$.
Accordingly, the resultant different output currents of the DD detector
can distinguish the qubit states.

\section*{ ``$n$"-Resolved Master Equation}

In the reduced description, the central dots of the detector and
the qubit are the {\it system of interest}, and the two reservoirs
of the detector are the {\it environment}. The first step is to
derive a master equation for the system of interest. Moreover, in
order to relate the master equation also to the output of the
detector, one should obtain a ``$n$"-resolved master equation.
Here, ``$n$" denotes the number of electrons in certain specified
time interval that have tunneled through the left or right
junction of the transport device. Following the previous work
about the master equation
\cite{Shn98,Gur96,Gur05,Gur06,Li051,Li052}, we obtain
\begin{align}\label{Eq3}
 \dot{\rho}^{(n_R)}=&-i\mathcal{L}\rho^{(n_R)}-\frac{1}{2}\{[a_c^\dagger,
 A_c^{(-)}\rho^{(n_R)}-\rho^{(n_R)}A_c^{(+)}]\nl&+a_d^\dagger A_d^{(-)}\rho^{(n_R)}
 +\rho^{(n_R)}A_d^{(+)}a_d^\dagger
 \nl&-[a_d^\dagger\rho^{(n_R+1)}A_{d}^{(+)}+A_{d}^{(-)}
 \rho^{(n_R-1)}a_d^\dagger]+ {\rm H.c.}  \} .
\end{align}
Note that throughout this paper we shall use the unit system of $\hbar=e=k_B=1$.
Shown above is in fact the ``$n_{R}$"-resolved master equation, with ``$n_{R}$"
the number of electrons tunneled through the right junction.
Similar equation can be carried out for the left-junction specified
tunneled electrons.
The superoperators in \Eq{Eq3} read
$A_\alpha^{(\pm)}=C_\alpha^{(\pm)}(\pm\mathcal{L})a_\alpha$.
$C_\alpha^{(\pm)}(\pm\mathcal{L})$ are the spectral functions of the two
reservoirs, which are the the Fourier transform of the correlation
functions, i.e.,
$C_\alpha^{(\pm)}(\pm\mathcal{L})=\int_{-\infty}^{+\infty}dt
C_\alpha^{(\pm)}(t)e^{\pm i\mathcal{L}t}$,
with $C_\alpha^{(+)}(t)=\langle f_\alpha^{\dg}(t)f_\alpha\rangle$
and $C_\alpha^{(-)}(t)=\langle f_\alpha(t)f_\alpha^{\dg}\rangle$.

Note that the Liouvillian $\mathcal{L}$ is defined by
$\mathcal{L}(\cdots)=[H_S,\cdots]$.
To explicitly carry out the action of its arbitrary function on
an operator (e.g. $a_c$ or $a_d$), a convenient way is doing it
in the eigenstate basis of $H_S$.
In this basis, the matrix element of the arbitrary function of
$\mathcal{L}$ is obtained by simply replacing $\mathcal{L}$ with the
energy difference of the two basis states.

\section*{Readout Characteristics}
Note that $\rho^{(n)}$ contains rich information about the measurement.
From it, one can obtain the distribution function of
the tunneled electron numbers, the output current and the noise spectrum.
Quite clearly,
the distribution function reads $P(n_R,t)={\rm Tr}[\rho^{n_R}(t)]$,
where the trace is over the states of the system of interest.
Then, the current through the right junction is
\begin{align}\label{IRt}
 I_R(t)=&\sum_{n_R}{\rm Tr}\{n_R\dot{\rho}^{(n_R)}\}\nl=&\frac{1}{2}
     {\rm Tr} \{[a_d^\dagger
 A_d^{(-)}-A_d^{(+)}a_d^\dagger]
 \rho(t)+{\rm H.c.}\},
\end{align}
where $\rho(t)=\sum_{n_R} \rho^{(n_R)}(t)$.
$\rho(t)$ satisfies the usual {\it unconditional} master equation,
which can be straightforwardly
obtained in this context by summing up \Eq{Eq3} over ``$n_R$".
Similar result as \Eq{IRt} can be obtained for $I_L(t)$,
the current through the left junction.

Now we formulate the calculation of the output power spectrum.
It is well known that the noise spectrum is a measure of the temporal
correlation of the current.
The temporal fluctuating currents through the left and right junctions,
even in steady state, are not equal to each other.
The circuit current, which is typically the measured quantity in most
experiments, is a superposition of the left and right currents, i.e.,
$I(t)=aI_L(t)+bI_R(t)$.
Here the coefficients $a$ and $b$ satisfy $a+b=1$, and depend on the
junction capacitances of the detector \cite{Butti}.
Note that this capacitive geometry is {\it not} necessarily in accordance
with the tunnel couplings. For very asymmetric tunnel couplings, the
capacitive geometry can be quite symmetric.
In what follows we shall see that this is in fact the setup we want to suggest.

In view of the charge conservation, i.e., $I_L=I_R+\dot{Q}$,
where $Q$ is the charge on the central dots, we obtain
$ I(t)I(0)=aI_L(t)I_L(0)+bI_R(t)I_R(0)-ab\dot{Q}(t)\dot{Q}(0)$.
Accordingly, the noise spectrum is a sum of three parts
\begin{eqnarray}\label{Sc}
S(\omega)=a S_L(\omega)+bS_R(\omega)-ab\omega^2S_Q(\omega),
\end{eqnarray}
where $S_{L/R}(\omega)$ is the noise spectrum of the current through
the left (right) junction,
and $S_Q(\omega)$ characterizes the charge fluctuations on the central dots.
For $S_{L/R}(\omega)$, it follows the MacDonald's formula
\begin{eqnarray}\label{SW-1}
S_{\alpha}(\omega)=2\omega\int_0^{\infty}dt {\rm sin}\omega t
\frac{d}{dt}\langle n_{\alpha}^2(t)\rangle
\end{eqnarray}
where $\langle
n_{\alpha}^2(t)\rangle$=$\Sigma_{n_{\alpha}}n_{\alpha}^2
{\rm Tr}\rho^{(n_{\alpha})}(t)$=$\Sigma_{n_{\alpha}}n_{\alpha}^2P(n_{\alpha},t)$.
With the help of \Eq{Eq3}, we further obtain
\begin{eqnarray}\label{dn2t}
\frac{d}{dt}\langle n_{\alpha}^2(t)\rangle
={\rm Tr} [2\mathcal{J}_{\alpha}^{(-)}N^{^\alpha}(t)
+\mathcal{J}_{\alpha}^{(+)}\rho+ {\rm H.c.}],
\end{eqnarray}
where the {\it particle-number} matrix reads
$N_{\alpha}(t)\equiv\sum_{n_{\alpha}}n_{\alpha}\rho^{(n_{\alpha})}(t)$,
and the superoperator means
\begin{align}\mathcal{J}_{\alpha}^{(\pm)}(\cdots)
=&\frac{1}{2}[A_{\mu}^{(-)}(\cdots)a_{\mu}^{+}\pm
a_{\mu}^{+}(\cdots)A_{\mu}^{(+)}] .
\end{align}
In this last equation, $\mu=c$ if $\alpha=L$; and
$\mu=d$ if $\alpha=R$.

Following Ref.\ \onlinecite{Li053}, it will be very convenient to work
in the frequency domain. Inserting \Eq{dn2t} into (\ref{SW-1}) we obtain
\begin{align}
S_{\alpha}(\omega)=&2\omega
{\rm Im}[{\rm Tr} \{2(\mathcal{J}_{\alpha}^{(-)}\tilde{N}^{\alpha}(\omega)
+[\mathcal{J}^{(-)}_{\alpha}\tilde{N}^{\alpha}(-\omega)]^{\dg})\nl
&+(\mathcal{J}_{\alpha}^{(+)}\tilde{\rho}(\omega)
+[\mathcal{J}_{\alpha}^{(+)}\tilde{\rho}(-\omega)]^{\dg})\}] ,
\end{align}
where
$\tilde{N}^{\alpha}(\omega)=\int_0^{\infty}dtN^{\alpha}(t)e^{i\omega t}$,
and $\tilde{\rho}(\omega)=\int_0^{\infty}dt\rho^{st}e^{i\omega t}$.
Note that $\rho^{st}$ is the stationary state density matrix,
which is time-independent.
We thus have $\tilde{\rho}(\omega)=i\rho^{st}/\omega$.
For $N^{\alpha}(\omega)$, it can be easily obtained by solving a set of
algebraic equations after Fourier-transforming the equation of motion of
$N^{\alpha}(t)$, as have been clearly described in Ref.\ \onlinecite{Li053}.

Concerning the charge fluctuations on the central dots, we define the
noise spectrum as
\bea
S_Q(\omega)&=&\int_{-\infty}^{\infty}d\tau\langle
    N(\tau)N+NN(\tau)\rangle e^{i\omega\tau}   \nl
&=& 4 {\rm Re} \left[ \int_0^{\infty}d\tau
    S(\tau)e^{i\omega\tau} \right],
\eea
where we have introduced $S(\tau)=\langle N(\tau)N\rangle$.
More explicitly, it can be expressed as
$S(\tau)= {\rm Tr Tr}_B[U^{\dg}(\tau)NU(\tau)N\rho^{st}\rho_B]$,
where $U(\tau)=e^{-iH\tau}$, and $N$ is the the electron number operator
of the central dots of the detector.
Using the cyclic property under trace,
we obtain $S(\tau)= {\rm Tr} [N\sigma(\tau)]$, and
$\sigma(\tau)\equiv {\rm Tr}_B[U(\tau)N\rho^{st}\rho_BU^{\dg}(\tau)]$.
Obviously, $\sigma(\tau)$ satisfies the same equation of the
reduced density matrix $\rho(\tau)$. The only difference is
the initial condition, for $\sigma(\tau)$ which is $\sigma(0)=N\rho^{st}$.
Similar to the above, from the equation of motion of $\sigma(\tau)$,
its Fourier counterpart $\tilde{\sigma}(\omega)$ can be straightforwardly
carried out. Then, the charge fluctuation spectrum is obtained as
$S_Q(\omega)=4 {\rm Re Tr}[N\tilde{\sigma}(\omega)] $.

\begin{figure}[h]
\begin{center}
\includegraphics[width=9cm]{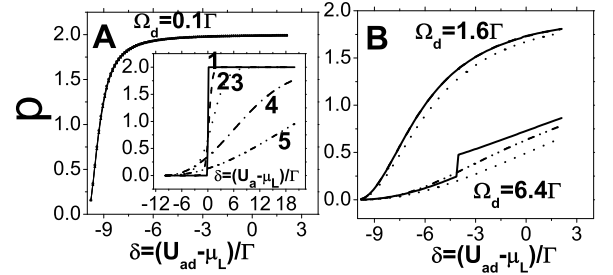}
\caption{\label{fig2} Tolerance of (not low-enough) finite
temperatures of the visibility, $p=|\Delta I|/\bar{I}=2|\Delta
I|/(I_a+I_b)$, for different inter-dot couplings of the DD
detector: (A) $\Omega_d/\Gamma=0.1$; (B) $\Omega_d/\Gamma=1.6$ and
$6.4$. 
As a comparison, the result of single-dot SET is plotted in the
inset of (A), where $U_{a}$ is the Coulomb interaction between the
qubit electron in dot ``$a$" and the transport electron in the
central dot of SET. 
Results are illustrated for three temperatures for the DD
detector, $T/\Gamma=0$, 1.6, and 12.8 (corresponding to the solid,
dot-dot-dashed, and dotted lines); while for five temperatures for
the single-dot SET, $T/\Gamma=0$, 0.4, 1.6, 6.4 and 12.8 (labeled
by ``1", ``2", $\cdots$, and ``5" ). For the DD detector, the
results are indistinguishable for small $\Omega_d$ as shown in
(A); the immunity against temperature will be weakened only for
large $\Omega_d$, as shown in (B). In contrast, the visibility of
the single-dot SET is affected by temperatures much more
sensitively.
In the whole calculations throughout the work, we assume that
$\Gamma_L=\Gamma$, and use $\Gamma$ as the energy unit. For the
result shown in this figure, we chose $\Gamma_R=\Gamma$. Other
parameters are adopted as: $E_c=E_d=0$, $U_{ac}=U_{bc}=U_{bd}=0$,
$\mu_L=10\Gamma$, and $\mu_R=-10\Gamma$.  }
\end{center}
\end{figure}

Based upon the above formalism, we now investigate the readout
characteristics of the DD detector.
The first important quantity to characterize the detector
is the {\it visibility}, which is defined by
$p=|\Delta I|/\bar{I}=2|I_a-I_b|/(I_a+I_b)$. In Fig.\ 2 we plot
the visibility against the qubit-detector interaction strength
``$U_{ad}$", by taking the temperature and the dot-dot coupling
strength ``$\Omega_d$" of the DD detector as other comparative
parameters.
By comparing the results shown in Fig. 2(a) and (b), it is found
that for smaller $\Omega_d$ the visibility can more easily
approach the ideal value of 2, by increasing the interaction
strength ``$U_{ad}$". In practice, controlling $U_{ad}$ is
difficult. However, engineering $\Omega_d$ is relatively easy,
which opens a way to enhance the visibility as revealed in Fig.\
2.
In this context, one should also notice another major advantage of
the DD detector, say, its better tolerance to finite temperatures.
From Fig.\ 2 we see that the finite temperature does not
sensitively affect the operation of the DD detector under proper
parametric conditions, particularly for small $\Omega_d$ as shown
in Fig.\ 2(a). Contrary to that, in the inset of Fig.\ 2(a), the
result of single-dot detector is presented, of which the
visibility sensitively depends on the temperature.
All these features can be easily understood in terms of resonant
tunnelling through the double dots and single dot, respectively.

\begin{figure}
\begin{center}
\includegraphics[width=9cm]{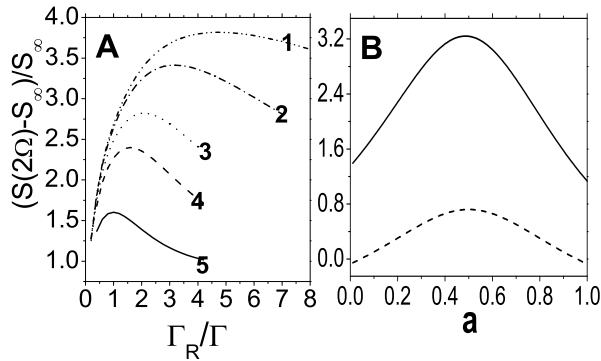}
\caption{\label{fig3} Configuration symmetry dependence of the
peak-to-pedestal ratio of the output power spectrum: (A)
$\Gamma_R$-dependence, and (B) capacitive coupling dependence.
Note that in (B) the parameter $a$ characterizes the capacitive
coupling symmetry (see the main text for its more detailed
explanation).
The major parameters are the same as in Fig.\ \ref{fig2}, except
for the differences as follows: 
In present result, it is assumed that $\Omega=\Omega_d=0.2\Gamma$,
and the temperature $T=0$.
In (A), we assume a symmetric configuration of capacitive
coupling, i.e., $a=1/2$; and assume the Coulomb interaction
strengths as (1) $U_{ad}/\Gamma=12$, (2) $U_{ad}/\Gamma=6$, (3)
$U_{ad}/\Gamma=3$, (4) $U_{ad}/\Gamma=2$, and (5)
$U_{ad}/\Gamma=1$.
In (B), in addition to the result depicted by the solid curve,
which corresponds to the suggested location of the qubit nearby
the {\it right dot} of the DD detector (as schematically shown in
Fig.\ 1), we also plot the result by the dashed curve for the
result of configuration with the qubit nearby the {\it left dot}.
For the former configuration, $U_{ac}=0$ and $U_{ad}=6\Gamma$;
while for the latter, $U_{ad}=0$ and $U_{ac}=6\Gamma$. And for
both configurations, $\Gamma_R=2\Gamma$ is commonly used. }
\end{center}
\end{figure}

In addition to the visibility, the quality of a quantum detector
is well characterized by the signal-to-noise ratio, i.e.,
the {\it peak-to-pedestal} ratio of the output power spectrum. 
Not as in Ref.\ \onlinecite{Gur06}, where the capacitively
asymmetric coupling model, i.e., with $a=0$ and $b=1$,  was taken
into account, below we calculate the noise spectrum in general
under arbitrary capacitive couplings. And in particular, the
symmetric coupling, say, $a=b=1/2$, will be focused.
Notably, from Fig.\ 3(a) we find that the peak-to-pedestal ratio
is sensitively affected by the tunnel rate $\Gamma_R$ of the right
junction, where the measured qubit is placed nearby. This feature
is in qualitative agreement with that found by Gurvitz {\it et al}
\cite{Gur06}, although a different definition of the signal-to-noise
ratio was employed there.

In Fig.\ 3(b) we show the effect of the capacitive coupling
symmetry. It is found that the signal-to-noise ratio will reach
the maximum at the symmetric coupling, i.e., when $a=1/2$. 
This is because the charge number fluctuation on the two dots of
the detector has negative contribution to the noise spectrum,
thus largely suppresses the background noise. As a
consequence, the peak-to-pedestal ratio is enhanced for more
symmetric coupling, and reaches the maximum at $a=1/2$.
In Fig.\ 3(b), the solid (dashed) curve corresponds to the result
of the measured qubit next to the right (left) dot of the DD
detector. This remarkable difference reflects another interesting
symmetry effect of the setup configuration.

\begin{figure}
\begin{center}
\includegraphics[width=8cm]{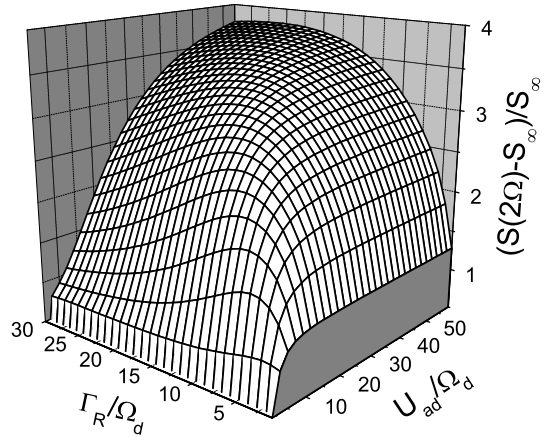}
\caption{\label{fig4} 3D plot of the peak-to-pedestal ratio of the
output power spectrum as a function of $\Gamma_R$ and $U_{ad}$.
Relevant parameters are the same as in Fig.\ \ref{fig3}A.}
\end{center}
\end{figure}

Notice that $\Gamma_R$ and $\Omega_d$ are two controllable
parameters in practice. We thus re-plot the signal-to-noise ratio
versus the scaled $\Gamma_R$ and $U_{ad}$ by $\Omega_d$, in order
to gain the entire landscape more clearly as shown in Fig.\ 4. 
In this context, we remark that the peak-to-pedestal ratio of the
DD detector can approach the upper limit of 4 of the {\it ideal}
QPC detector \cite{Korotkov01}, under proper parametric conditions
as indicated by Fig.\ 4. 
This conclusion is in contrast with that by Gurvitz {\it et al}
\cite{Gur05,Gur06}. There, it was concluded that both the
single-dot and double-dots detectors are only {\it sensitive}
measurement devices (i.e. with desirable visibility), but {\it
cannot reach the effectiveness of an ideal QPC detector}.
By tilting the tunnel coupling such that $\Gamma_R>>\Gamma_L$,
Gurvitz {\it et al} found that the signal-to-noise ratio can be
considerably enhanced. However, their calculation was restricted
to the {\it capacitively} asymmetric coupling model, i.e., with
$a=0$ and $b=1$. In this case, the upper limit of the
signal-to-pedestal ratio is 2. Here, as clearly shown by Fig.\
3(b), our calculation shows that under the symmetric condition
$a=b=1/2$ the signal-to-pedestal ratio is maximal, and can in
principle approach the value of 4, which is the upper limit of the
ideal QPC detector \cite{Korotkov01}.

As a brief summary, in the above we revealed three types of configuration
dependence:
(i) left-versus-right location of qubit with respect to
the quantum dots of the DD detector,
(ii) relative coupling to the right electrode (i.e. $\Gamma_R$-dependence),
and (iii) capacitive coupling.
While (iii) was resolved in terms of the role of
the central charge-number fluctuations in Eq.\ (4),
we would like to elaborate on (i) and (ii) further as follows.

If the qubit is next to the right dot of the DD detector, the interacting time
is relatively shorter than the one when the qubit locates nearby the left dot.
As a result, smaller back-action induced dephasing rate
is anticipated from general consideration, which in turn results
in the larger signal-to-noise ratio.
Similar reasoning can partially apply to the $\Gamma_R$-dependence in Fig.\ 3(a).
However,  in addition to the interacting time, the current through the detector,
which influences the interacting strength, would also affect the
back-action dephasing.
For the DD detector, the current difference associated with
different qubit states, which is nothing but the \emph{signal} ,
shows a turnover behavior with maximum at
$\Gamma_R=2\sqrt{2}\Omega_d$. Explicit expression is referred to
Eq.\ (4.4) in Ref.\ \onlinecite{Gur97}, see also Eq.\ (11) in next
section of the present work. The dephasing rate shown in Fig.\
5(a) largely follows the behavior of the signal current. 
With both the dephasing rate and the signal current in mind,
the $\Gamma_R$-dependence in Fig.\ 3(a) can be accordingly understood.
Note that the small back-action dephasing tends to enhance the signal-to-noise
ratio, while in contrast the small signal current would reduce it.
The particular line-shape of the signal-to-noise ratio versus the $\Gamma_R$
is thus a result of these two competing effects,
which lead to its turnover behavior and the optimal $\Gamma_R$ differing
from the dephasing rate in Fig.\ 5(a).

\section*{Dephasing and Measurement Times }

In the orthodox Copenhagen postulate for quantum measurement, the
measured wavefunction collapses onto one of the eigenstates of the
observable {\it instantaneously}. In contrast to that, the
wavefunction collapse in real device must need some time, i.e.,
the {\it measurement time}.
On the other hand, during the collapsing process,
dephasing between the superposed wavefunction components must take place
{\it before} reading out the result.
Therefore, the ratio of the dephasing time to the measurement time
is another deep criterion to
characterize the {\it efficiency of quantum measurement}.
In the following we carry out a quantitative analysis for the DD detector.

The analysis is also based on the ``$n$"-resolved master equation.
Since in this context we are interested in the measurement-induced collapse
of wavefunction, we consider the measurement of the idle state of the qubit.
We thus set $\Omega=0$, i.e., switch off the qubit state oscillation.
Accordingly, all the mixing terms, i.e., those proportional
to $\Omega$, disappear in \Eq{Eq3}.
And the density matrix of the system factorizes into three independent groups.
Furthermore, we restrict our analysis to zero temperature,
and assume that $E_i=0$ for $i=a,b,c$, and $d$, and $U_{ac}=U_{bc}=U_{bd}=0$.
By Fourier transforming the resultant ``$n$"-resolved master
equation, i.e., defining
$\rho(k,t)=\sum_{n_R}\rho^{(n_R)}(t)e^{i{n_R}k}$, we obtain
\begin{subequations}
\begin{align}
\dot{\rho}_{aa}^{00}=&-\Gamma_L\rho_{aa}^{00}+\Gamma_Re^{ik}\rho_{aa}^{22}\\
\dot{\rho}_{aa}^{11}=&-i\Omega_d[\rho_{aa}^{21}-\rho_{aa}^{12}]
+\Gamma_L\rho_{aa}^{00}\\
\dot{\rho}_{aa}^{22}=&-i\Omega_d[\rho_{aa}^{12}-\rho_{aa}^{21}]
-\Gamma_R\rho_{aa}^{22}\\
\dot{\rho}_{aa}^{12}=&iU_{ad}\rho_{aa}^{12}-i\Omega_d[\rho_{aa}^{22}
-\rho_{aa}^{11}]-\frac{1}{2}\Gamma_R\rho_{aa}^{12}\\
\dot{\rho}_{bb}^{00}=&-\Gamma_L\rho_{bb}^{00}+\Gamma_Re^{ik}\rho_{bb}^{22}\\
\dot{\rho}_{bb}^{11}=&-i\Omega_d[\rho_{bb}^{21}-\rho_{bb}^{12}]
+\Gamma_L\rho_{bb}^{00}\\
\dot{\rho}_{bb}^{22}=&-i\Omega_d[\rho_{bb}^{12}-\rho_{bb}^{21}]
-\Gamma_R\rho_{bb}^{22}\\
\dot{\rho}_{bb}^{12}=&-i\Omega_d[\rho_{bb}^{22}-\rho_{bb}^{11}]
-\frac{1}{2}\Gamma_R\rho_{bb}^{12}\\
\dot{\rho}_{ab}^{00}=&-\Gamma_L\rho_{ab}^{00}+\Gamma_Re^{ik}\rho_{ab}^{22}\\
\dot{\rho}_{ab}^{11}=&-i\Omega_d[\rho_{ab}^{21}-\rho_{ab}^{12}]
+\Gamma_L\rho_{ab}^{00}\\
\dot{\rho}_{ab}^{22}=&-iU_{ad}\rho_{ab}^{22}-i\Omega_d[\rho_{ab}^{12}
-\rho_{ab}^{21}]-\Gamma_R\rho_{ab}^{22}\\
\dot{\rho}_{ab}^{12}=&-i\Omega_d[\rho_{ab}^{22}-\rho_{ab}^{11}]
-\frac{1}{2}\Gamma_R\rho_{ab}^{12}\\
\dot{\rho}_{ab}^{21}=&-iU_{ad}\rho_{ab}^{21}-i\Omega_d[\rho_{ab}^{11}
-\rho_{ab}^{22}]-\frac{1}{2}\Gamma_R\rho_{ab}^{21}
\end{align}\label{Eq6}
\end{subequations}
We see that these equations split into three groups, i.e.,
(\ref{Eq6}a)-(\ref{Eq6}d), (\ref{Eq6}e)-(\ref{Eq6}h), and
(\ref{Eq6}i-\ref{Eq6}m).
Here, the density matrix elements
$\rho_{mn}^{ij}=\langle im|\rho|jn\rangle$.
$|i\rangle$ and $|j\rangle$ denote the occupation states of the DD detector,
i.e., $|0\rangle\equiv |00\rangle $,
$|1\rangle\equiv |10\rangle$ and $|2\rangle\equiv |01\rangle$, respectively;
while $|m\rangle$ and $|n\rangle$ denote the qubit states $|a\rangle$ and $|b\rangle$.

We now consider the characteristic solutions of the the above three groups
of equations, i.e., solutions proportional to $e^{i\omega t}$.
Technically, for each group of \Eqs{Eq6}, we can obtain five eigenvalues.
For small values of $k\ll 1$, from the former two groups of \Eqs{Eq6}
we obtain the smallest two eigenvalues
$\omega_j(k)=(k+\frac{1}{2}if^jk^2)\Gamma^j$, with $j=a$ and $b$, respectively,
which are most relevant to present analysis.
$\Gamma^j$ are the wave-packet's group velocities,
which actually correspond to the stationary currents $I_j$,
with respect to the qubit in state $|j\ra$;
$f^j$ are the respective Fano factors.
Explicitly, from Eqs.\ (\ref{Eq6}a)-(\ref{Eq6}h),
$\Gamma^j$ and $f^j$ are obtained as
\begin{eqnarray}
\Gamma^j=\frac{\Omega_d^2\Gamma_R}{(\frac{\Gamma_R^2}{4}+U_j^2)
+\Omega_d^2\Gamma_R(\frac{1}{\Gamma_L}+\frac{2}{\Gamma_R})}
\equiv\frac{\Omega_d^2\Gamma_R}{A_j}
\end{eqnarray}
\begin{eqnarray}
f^j=1+\frac{2\Omega_d^2}{A_j} \left[ 2 - \frac{\Gamma_R^2+
  (1+\frac{\Gamma_R}{\Gamma_L}) (\frac{\Gamma_R^2}{4}+U_j^2+4\Omega_d^2) }
  {A_j} \right]. \nl
\end{eqnarray}
Here, the Coulomb interaction energy $U_a=U_{ad}$, and $U_b=0$,
with the convention that the two dot-states of the DD detector are in resonance
if the qubit is in state $|b\ra$, and in off-resonance by an energy $U_a=U_{ad}$
if the qubit is in state $|a\ra$.
Quantitatively, the measurement time can be defined as the required time for
signal-to-noise ratio approaching unity.
This condition leads to \cite{Sch01,Ale97}
\begin{eqnarray}
t_{m}=\left( \frac{\sqrt{2f^a\Gamma^a}
+\sqrt{2f^b\Gamma^b}}{\Gamma^a-\Gamma^b} \right)^2  ~.
\end{eqnarray}

The dephasing time can be obtained by analyzing
Eqs.\ (\ref{Eq6}i)-(\ref{Eq6}m) for $k=0$.
Similarly, solve the (five) eigenvalues $\lambda_i$ of these equations,
then determine the dephasing time in terms of
$t_d={\rm max {[Im}\lambda_i]^{-1}}$.
Importantly, the quantum measurement efficiency is obtained via
$\eta=1/(2\Gamma_d t_m)$, where $\Gamma_d=1/t_d$.

\begin{figure}
\begin{center}
\includegraphics[width=8cm]{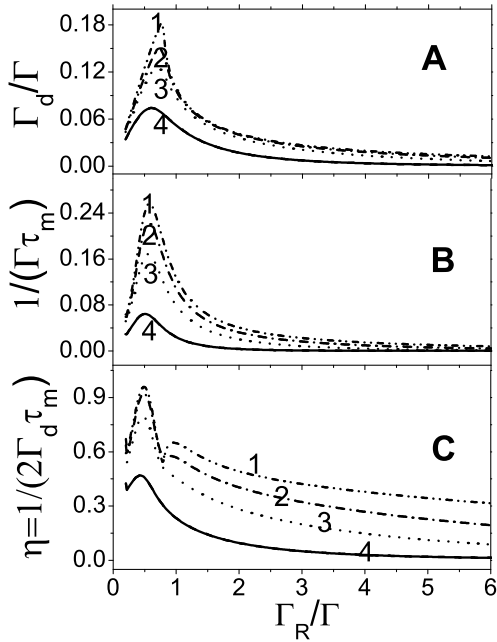}
\caption{\label{fig5} $\Gamma_R$-dependence (i.e. asymmetric
effect) of the dephasing rate (A), the measurement time (B), and
the quantum measurement efficiency (C). Coulomb interaction
strengths: (1) $U_{ad}/\Gamma=12$, (2) $U_{ad}/\Gamma=6$, (3)
$U_{ad}/\Gamma=3$, and (4) $U_{ad}/\Gamma=1$. Other relevant
parameters are referred to Fig.\   \ref{fig3}A.}
\end{center}
\end{figure}

In Fig.\ 5 we plot the $\Gamma_R$-dependence of the measurement time,
dephasing rate, and the quantum efficiency of measurement.
At the end of the previous section, we have explained the
$\Gamma_R$-dependence of the signal-to-noise ratio
in terms of dephasing rate and signal current,
and pointed out that the dephasing rate is roughly proportional
to the signal current, which is now depicted in Fig.\ 5(a).
From the general viewpoint of quantum measurement,
the measurement rate, i.e., the rate of information gain,
should follow the back-action dephasing rate.
This is shown in Fig.\ 5(b).

The quantum measurement efficiency, which is
the ratio of the dephasing time and the measurement time,
is shown in Fig.\ 5(c).
We notice that it \emph{does not} well match the behavior of
the signal-to-noise ratio in Fig.\ 3(a),
although both have maxima at proper (different) $\Gamma_R$.
This feature is not surprising, since the quantum measurement
efficiency is anyhow an alternative criterion to qualify the
measurement process. That is, it describes how fast the
information is gained against the back-action dephasing \cite{Clerk03}.

Remarkably, in contrast with the usual statement that the
single-electron-transistor is {\it not} an ideal detector
\cite{Dev00,Sch01}, it is found here that the double-dot
SET can approximately reach the quantum limit of efficiency
under appropriate parametric conditions, see Fig.\ 5(c).
However, these parametric conditions do not simultaneously
promise the maximal signal-to-noise ratio.
It is noticed that deep work by Clerk \emph{et al} had focused on
the measurement efficiency of quantum scattering detectors
\cite{Clerk03,Clerk04}. Using the scattering matrix
formalism, general conditions for quantum limited measurements
were carried out. Unfortunately, it is not convenient, if not
possible, to apply the scattering matrix formalism to the SET-type detectors.
Following the line of Clerk \emph{et al}, especially using the concept
of information gain and loss,
further elaboration on the quantum limit of efficiency found here
is interesting and an open question for future work.

\section*{Conclusion and Discussions}

To summarize, we have presented a study for the quantum measurement
characteristics of double-dot SET.
The study was based on a full analysis of the setup configuration geometries ,
i.e., in terms of the tunneling strengths, capacitive couplings,
and the location of the qubit with respect to the DD detector.
We found that the DD detector can approach the signal-to-noise ratio
of an ideal QPC detector, provided the symmetric capacitive coupling
is taken into account.
The measurement time, the back-action dephasing time
and the measurement efficiency were calculated.
It was found that the quantum limit of efficiency
can be reached under proper parametric conditions, although
which differ from the ones for obtaining
the maximal signal-to-noise ratio.


Finally, we make a few remarks on issues relevant to the present work.
In ref.\ \onlinecite{Clerk02}, the measurement properties
of the superconducting SET (SSET) were analyzed, where both the
coherent cooper-pair tunneling and the quasi-particle tunneling
were taken into account to contribute the measurement current. It
was concluded that the Cooper-pair resonance process allows for a
much better measurement than a similar nonsuperconducting SET, and
can approach the quantum limit of efficiency under proper
parametric conditions.
In our opinion, the advantages of the SSET are largely a consequence of the
coherent tunneling of cooper pairs, a unique nature of superconductors.

About the nonsuperconducting SET, such as our semiconductor DD detector,
we do not expect that higher order tunnel processes can considerably
influence or improve the measurement efficiency.
Higher order (e.g. cotunneling) contribution, which leads to small
detection current, was calculated in the Coulomb-blockade regime
of SSET \cite{Kin03}, and was shown to have minor effect
on the measurement effectiveness, say, the signal-to-noise ratio.

It has come into our attention that the non-perturbative treatment
for strong qubit-detector coupling and arbitrarily strong
transmission detector has been recently an attractive research
subject \cite{Averin05,Naz07}. While in theses studies
the detector is a QPC, similar analysis for SET-type detectors
might be an interesting subject of future work.
However, typical experiments such as those performed by Marcus
group did not imply strong couplings of the double QDs with the
transport electrodes \cite{Mar04,Mar05}. In these
experiments, the charge configurations of coupled QDs were probed
by techniques such as the nearby QPC or direct transport
spectroscopy. In order to make the charge-states of the coupled
QDs well defined, the couplings of the double QDs with the
external (transport) electrodes should be relatively weak.

\vspace{5ex} {\it Acknowledgments.}
We thank professor S. A. Gurvitz for helpful discussion,
which drew our attention on the double-dot detector
and initiated the present study.
This work was supported by the National Natural Science
Foundation of China under grants No.\ 60425412 and No.\ 90503013,
the Major State Basic Research Project under grant No.2006CB921201,
and the Research Grants Council of the Hong Kong Government.



\clearpage





\end{document}